\documentclass[equations,12pt]{article}
\usepackage{graphicx,epsfig}
\input epsf
\def\LAMBDABAR {\hbox{$\lambda$\kern-0.52em\raise+0.45ex\hbox{--}\kern+0.2em}}
\def\rsim{>\kern-2.5ex\lower0.85ex\hbox{$\sim$}\ }
\def\lsim{<\kern-2.5ex\lower0.85ex\hbox{$\sim$}\ }

\begin{document}

\centerline{\bf Comment on ``Experimental Observation of Optical}
\centerline{\bf Rotation Generated in Vacuum by a Magnetic Field''}
\bigskip
\centerline{Adrian C. Melissinos}
 \centerline{\it Department of Physics
and Astronomy, University of Rochester, Rochester, NY  14627, USA}
\centerline{February 12, 2007}
\bigskip

In a recent letter [1] the PVLAS collaboration reported the
observation of optical rotation induced in vacuum by an external
magnetic field. More recently [2] the same group reported also the
observation of vaccuum birefringence under similar conditions.  In
this comment I wish to point out that the reported results are
excluded at the 95\% confidence level by a previous experiment [3]
carried out by the BRFT collaboration (Brookhaven, Rochester,
Fermilab, Trieste). In Table 1 we summarize the relevant parameters
of the two experiments.
\bigskip

\centerline{\bf Table 1. Comparison of the PVLAS and BRFT
experiments}
$$\begin{array} {llrll}
& {\rm PVLAS\ [1,2]} & {\rm BRFT\ [3]} \\

B^{2}_{e}\ (${\rm Effective magnetic field)}$  & 30 & 4\qquad {\rm T^{2}}  \\

\ell\ {\rm (Length\ of\ magnetic\ field\ region)} & 1 & 8.8 \qquad  {\rm m} \\

\hbar\omega\ {\rm(Photon \ energy)} & 1.16,\ 2.32 & 2.4 \qquad {\rm eV} \\
 \epsilon_{T}\ {\rm(Measured\ optical\ rotation)} & [170\pm 22]\times 10^{-9}\ &
0.6\times 10^{-9}\ {\rm rad} \\
&  & 95\% {\rm upper\ limit}\\

N_{R}\ {\rm (Number\ of\ passes)} & 44\times 10^{3} & 254 \\
 \psi_{T}\ {\rm (Measured\ ellipticity)} & [440 \pm ??] \times 10^{-9}\ & 51\times
10^{-9}\ {\rm rad} \\
&  & 95\% {\rm upper\ limit }\\
 N_{E}\ {\rm(Number\ of\ passes)} & 44\times 10^{3}(?) & 578

\end{array}$$

The two experiments are conceptually similar with the following
differences:  (1)  PVLAS uses a Fabry-Perot cavity to effect
multiple reflections while BRFT used an optical delay line (2) PVLAS
uses a fixed field magnet which is mechanically rotated.  In the
BRFT experiment the magnetic field was modulated by varying the
excitation current. The ellipticity induced by QED effects [4]
appears at a level $\sim 10^{4}$ times below the noise level of
either experiment.

The PVLAS group attributes their signals to the existence of a
scalar or pseudoscalar particle that couples to two photons [5]. The
observed rotation/pass, $\epsilon$, and ellipticity/pass, $\psi$,
are related to the coupling constant $g_{a\gamma\gamma} = 1/M$ and
to $m_a$, the mass of the scalar/pseudoscalar through [6]
\begin{eqnarray*}
\epsilon = \left({1\over M}\right)^{2} {B^{2}\omega^{2}\over
m_{a}^{4}} \sin^{2} \left({m_{a}^{2}\ell \over 4\omega}\right)
\end{eqnarray*}

\begin{eqnarray*}
\psi = \left({1\over M}\right)^{2} {B^{2}\omega^{2}\over m_{a}^{4}}
{1\over 2} \left[ {m_{a}^{2}\ell \over 2\omega} - \sin \left(
{m_{a}^{2}\ell \over 2\omega} \right)\right]
\end{eqnarray*}
In the above equations we have set $\hbar = c=1.$ We see that the
limits on the value of the inverse coupling, deduced from the
measured rotation and ellipticity, depend on the mass of the
scalar/pseudoscalar. This is because for a given path length massive
particles dephase (with respect to the laser beam) faster. The phase
advance is $2\pi(\ell{m_a}^2/2\omega)$ and beyond this point the
observable effects oscillate rapidly as a function of $m_a$, as can
be seen in the figure.

In Fig. 1 we show the values of $M$ (in GeV) as deduced from the two
experiments.

\medskip

\begin{tabular}{lr}
     PVLAS (from rotation) & red \\
     BRFT (from rotation 95\% confidence lower limit) & cyan \\
     PVLAS (from ellipticity) & green \\
     BRFT (from ellipticity 95\% confidence lower limit) & blue
\end{tabular}
\medskip

\noindent The regions {\it below} the cyan and blue curves are
excluded by the BRFT experiment, namely the PVLAS results are
excluded for $m_{a} < 0.7 \times 10^{-3}$ eV. The most recent
measurement of ellipticity by PVLAS coupled with their previous
measurement of optical rotation  fixes the mass and coupling at the
intersection of the red and green curves, namely at $m_a = 1.3\times
10^{-3}$ eV and $M = 3\times 10^{5}$ GeV. This point falls on the
enveloppe of the 95\% exclusion limit of the BRFT experiment.

As can be seen in Fig. 1 the values of $(1/M, m_a)$ measured by
PVLAS are in the region where their sensitivity is decreasing and
where the BRFT sensitivity is already oscillating rapidly (because
of the longer magnetic region). It is rather improbable that Nature
has conspired to place the scalar/pseudoscalar mass and coupling
exactly at the limit of the detectable phase space of both
experiments.

That the PVLAS effect could be due to instrumental background is
suggested from Fig. 2. Here a typical spectrum of the PVLAS
experiment is reproduced from their publication [1]. The signal
appears at the second harmonic of the rotation frequency, and there
should be no signal at the first harmonic. Instead, a {\it spurious
signal} is present, 25 dbV ($\sim$ 18 times) stronger than the
signal that is interpreted as new physics. The PVLAS group gives
{\it no explanation} for this signal. It is clearly related to the
rotating magnet field; it is an artefact introduced by the
experimental setup. It follows that the same background could
generate the second harmonic signal as well, especially when it is
suppressed by an order of magnitude.

\medskip

\noindent {\bf References}
\medskip

\noindent[1]  E.Zavattini et al.(PVLAS Collaboration) Phys. Rev.
Letters {\bf 96}, 110406 (2006).

\noindent[2]  E.Zavattini et al.(PVLAS Collaboration) Paper
presented by U.Gastaldi at

ICHEP-06, Moscow, 26/7-2/8, 2006. Preprint INFN-LNL-214(2006).

\noindent[3]  R.Cameron et al. Phys. Rev. {\bf D47}, 3707 (1993).

\noindent[4]  See for instance, S.L.Adler, Ann. Phys. (NY) {\bf 67},
599 (1971)

\noindent[5]  L.Maiani, R.Petronzio and E.Zavattini Physics Letters
{\bf 175}, 359 (1986)

\noindent[6]  G.Raffelt and L.Stodolsky Phys. Rev. {\bf D37}, 1237
(1986).


\begin{tabular}{l}

\end{tabular}

\begin{figure}[h]
\centering
\includegraphics[width=0.8\columnwidth]{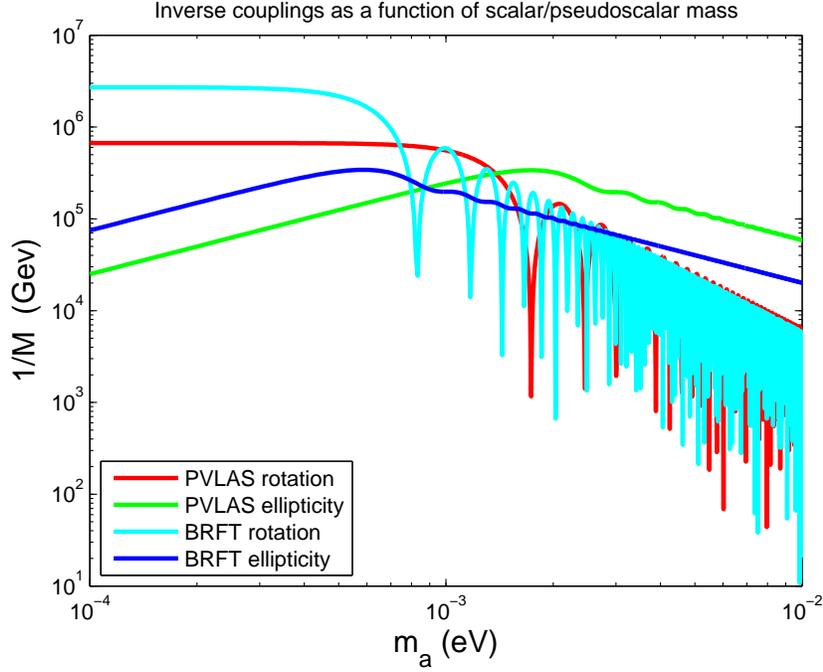}
\caption{The inverse coupling $M = 1/g_{\alpha\gamma\gamma}$ in GeV
as a function of the mass, $m_{\alpha}$ of the scalar/peudoscalar
particle in eV. (a) Cyan: BRFT limit from rotation (b) Blue: BRFT
limit from ellipticity (c) Red: PVLAS detection from rotation (d)
Green: PVLAS detection from ellipticity.}
\end{figure}

\begin{figure}[h]
\centering
\includegraphics[width=0.8\columnwidth, height = 2.5 in,keepaspectratio=false]{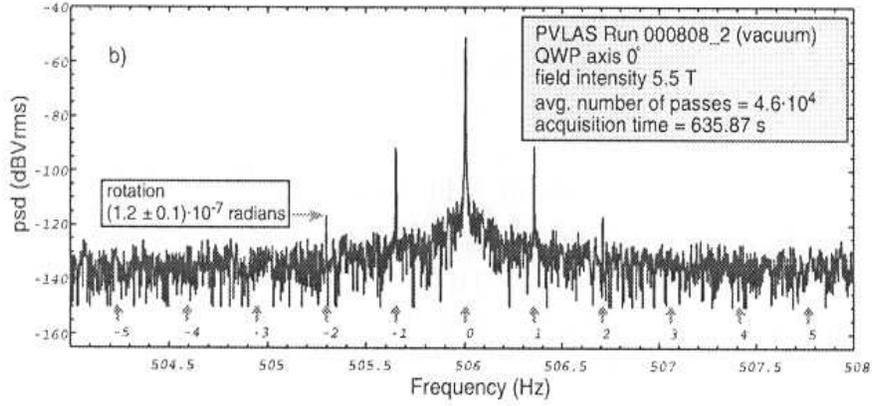}
\caption{The PVLAS Fourier spectrum of the observed rotation signal
(from [1]). The harmonics of the magnet rotation frequency are
labeled. The signal appears at the second harmonic, however the
first harmonic should be completely absent. The only possible
interpretation of the first harmonic is that it is due to
instrumental effects.}
\end{figure}

\end{document}